\documentclass[prl,twocolumn,showpacs,amsmath,amssymb]{revtex4}
\usepackage{graphicx}
\usepackage{enumerate}

\def \mate<#1|#2|#3>{\mbox{$\langle {#1}|\,{#2}\,|{#3}\rangle$}}

\begin{document}

\title{Equation of State for Nucleonic Matter and its Quark Mass Dependence\\
from the Nuclear Force in Lattice QCD}

\author{
Takashi Inoue$^{1}$,
Sinya Aoki$^{2,3}$,
Takumi Doi$^{4}$,
Tetsuo Hatsuda$^{4,5}$,\\
Yoichi Ikeda$^{4}$,
Noriyoshi Ishii$^{3}$,
Keiko Murano$^{4}$,
Hidekatsu Nemura$^{3}$,
Kenji Sasaki$^{3}$\\ (HAL QCD Collaboration)\\
}

\affiliation{
$^1${Nihon University, College of Bioresource Sciences, Kanagawa 252-0880, Japan}\\
$^2${Yukawa Institute for Theoretical Physics, Kyoto University, Kyoto 606-8502, Japan }\\
$^3${Center for Computational Sciences, University of Tsukuba, Ibaraki 305-8571, Japan}\\
$^4${Theoretical Research Division, Nishina Center, RIKEN, Saitama 351-0198, Japan}\\
$^5${Kavli IPMU (WPI), The University of Tokyo, Chiba 277-8583, Japan}
}

\begin{abstract}
Quark mass dependence of 
the equation of state (EOS) for nucleonic matter is investigated, 
on the basis of the Brueckner-Hartree-Fock method
with the nucleon-nucleon interaction extracted from lattice QCD simulations.
We observe  saturation of  nuclear matter at the lightest available quark mass 
corresponding to the pseudoscalar meson mass $\simeq 469$ MeV.  
Mass-radius relation of the neutron stars is also studied
with the EOS for neutron-star matter from the same nuclear force in lattice QCD. 
We observe that the EOS becomes stiffer  and thus 
the maximum mass of neutron star increases as
the quark mass decreases toward the physical point.
\end{abstract}
\maketitle

 The equation of state (EOS) for hadronic matter is a key quantity for understanding
 the physics of compact stars and  explosive phenomena in astrophysics.
 From the observational point of view,
 recent reports on massive neutron stars put stringent constraints on the EOS ~\cite{Demorest:2010bx}. 
 In the future, neutrinos from core-collapsed supernovae and gravitational waves from neutron star mergers
 will give further  constraints on the EOS~\cite{Sekiguchi:2011zd,Janka:2012sb}. 
 From the  theoretical point of view, various approaches to calculate the EOS have been pursued so far, e.g.
 the Brueckner-Bethe-Goldstone theory~\cite{Baldo:2011gz},
 the quantum Monte Carlo simulations~\cite{Carlson:2012mh},
 and the chiral effective theories~\cite{Hebeler:2013nza,Holt:2013fwa}. 
 Although they provide reasonable descriptions of the nuclear matter at low density,
 it is still beyond the scope of these approaches to answer the
 fundamental questions such as the quark-mass ($m_q$) dependence of 
 the nuclear saturation property and the  maximum mass of neutron stars.
 These questions can only be answered by 
 many-body techniques with hadronic interactions obtained by
 lattice QCD simulations for different $m_q$.
     
\begin{table}[t]
\caption{$M_{\rm PS}$, $M_{\rm V}$ and $M_{\rm B}$ denote hadron masses corresponding to
 the pseudoscalar meson, vector meson and octet baryon, respectively, with
 the $SU(3)$ symmetric hopping parameter $\kappa_{uds}$~\cite{Inoue:2011ai}. 
 Trajectory length $N_{\rm traj}$ and number of configurations $N_{\rm cfg}$ are also shown.}
\label{tbl:mass}
\smallskip
\centering
 \begin{tabular}{c|c|c|c|c}
   \hline  \hline
    ~~ $\kappa_{uds}$ ~~ & $M_{\rm PS}$ [MeV] &  $M_{\rm V}$ [MeV] &  $M_{\rm B}$ [MeV] &
   $N_{\rm cfg}\,/\,N_{\rm traj}$ \\
   \hline 
     0.13660 &   1170.9(7) &   1510.4(0.9) & 2274(2) & 420\,/\,4200 \\
     0.13710 &   1015.2(6) &   1360.6(1.1) & 2031(2) & 360\,/\,3600 \\
     0.13760 & ~\,836.5(5) &   1188.9(0.9) & 1749(1) & 480\,/\,4800 \\
     0.13800 & ~\,672.3(6) &   1027.6(1.0) & 1484(2) & 360\,/\,3600 \\
     0.13840 & ~\,468.6(7) & ~\,829.2(1.5) & 1161(2) & 720\,/\,3600 \\
   \hline  \hline
 \end{tabular}
\end{table}

 The purpose of this Letter is to make a first exploratory study for
 the nuclear and neutron matter EOS by 
 combining the Brueckner-Hartree-Fock (BHF) many-body theory with
 the nuclear force obtained from lattice QCD simulations. 
 In particular we study how the saturation develops in nuclear matter
 and how the mass-radius relation of the neutron star changes as a function of $m_q$:
 Such $m_q$ dependence of the EOS gives us useful information on the 
 physics of strongly interacting nucleons,
 even though the values of $m_q$ in this study are still away from the physical one.
 In addition, it is certainly important to establish a direct connection
 between  lattice QCD and the physics of the nucleonic matter.

 The nuclear force which we adopt in this Letter is taken from the
 zero-strangeness sector of the octet-baryon potentials
 in the flavor-$SU(3)$ limit calculated on the lattice~\cite{Inoue:2010es,Inoue:2011ai},
 where the renormalization group improved Iwasaki gauge action 
 and the nonperturbatively improved Wilson quark action were employed
 on a $32^3\times 32$ lattice with the lattice spacing $a=$0.121(2) fm.  
 The potentials were derived from the imaginary-time Nambu-Bethe-Salpeter wave functions
 by the HAL QCD method~\cite{Ishii:2006ec,HALQCD:2012aa,Aoki:2012tk}, at the
 quark masses corresponding to the pseudoscalar meson masses ($M_{\rm PS}$)
 ranging between 469 and 1171 MeV.
 Shown together in Table~\ref{tbl:mass} are 
 the vector meson mass ($M_{\rm V}$) and the baryon mass ($M_{\rm B}$) for these quark masses.

\begin{figure}[t]
\centering
\includegraphics[width=0.4\textwidth]{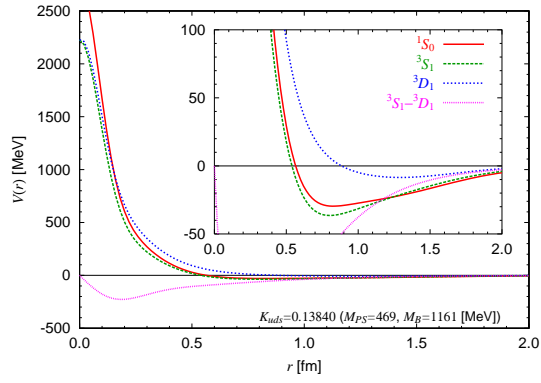}
\caption{The $N\!N$ potentials for S and D waves
extracted from lattice QCD data at $M_{\rm PS}$ = 469 MeV in the flavor-$SU(3)$ limit.
The lines are obtained by the least $\chi^2$ fit to the lattice data (see ~\cite{Inoue:2011ai} for the fit function).}
\label{fig:pot_K13840}
\end{figure}

\begin{figure}[t]
\centering
\includegraphics[width=0.4\textwidth]{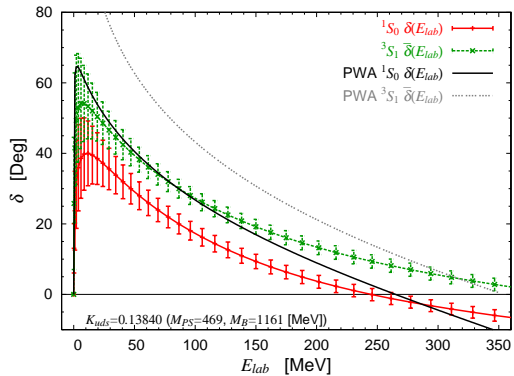}
\caption{
Phases shifts of $N\!N$ scattering as a function of energy in the laboratory frame,
extracted from lattice QCD data at $M_{\rm PS}$ = 469 MeV in the flavor-$SU(3)$ limit:
The black and gray dashed lines are partial wave analysis (PWA) of experimental data taken
from NN-OnLine (http:\slash\slash{}nn-online.org\slash) .}
\label{fig:ph_K13840}
\end{figure}

 In  Fig.~\ref{fig:pot_K13840}, we show the $N\!N$ potentials obtained from fits to the lattice data in S and D-waves at $M_{\rm PS}\simeq 469$ MeV.
 These potentials share common features with phenomenological potentials, i.e.,
 a strong repulsive core at short distance and the attractive pocket at intermediate distance,
 so that the $^1{\rm S}_0$ phase shift in  Fig.~\ref{fig:ph_K13840} 
 shows qualitatively similar behavior with the experimental data~\cite{Inoue:2011ai}.
 While the phase shift in the $^3{\rm S}_1$ channel
 shows stronger attraction at low energies than that of the  $^1{\rm S}_0$ channel due to
 $^3{\rm S}_1$-$^3{\rm D}_1$ mixing, it is still insufficient to form 
 the deuteron bound state even at  $M_{\rm PS}\simeq 469$ MeV~\cite{Inoue:2011ai,others}.
 Although we found no bound state in two and three nucleon systems,
 there exists a four-nucleon bound state, namely, $^4$He,
 with about 5 MeV binding energy  at $M_{\rm PS}\simeq 469$ MeV~\cite{Inoue:2011ai}.

Let us now study the EOS for nuclear matter and neutron matter
in the leading order of the Brueckner-Bethe-Goldstone (BBG) hole-line expansion, i.e.,  
the Brueckner-Hartree-Fock (BHF) theory (see e.g.~\cite{Baldo:2011gz}), where 
the total ground-state energy $E$
at zero temperature with the nucleon mass $M_N$ and the Fermi momentum $k_F$ reads
\begin{equation}
 E  = \sum_{k}^{k_F} \frac{k^2}{2 M_N} 
    + \frac{1}{2} \sum_{k,k'}^{k_F} {\rm Re\,} \langle k k'| G(e(k)+ e(k')) |k k' \rangle_A
\label{eqn:gene}
\end{equation}
with $|k k'\rangle_A \equiv |k k'\rangle - |k' k \rangle$.
 To simplify the notation, spin and isospin indices of the nucleons are included in the 
 label $k$.
The $G$ matrix, describing the in-medium effective interaction of the two nucleons,
obeys the Bethe-Goldstone integral equation with the bare interaction $V$, 
\begin{eqnarray}
 \!\!\!\!\!\!\!\!
  & & \langle k_1 k_2 |G(\omega)|  k_3 k_4 \rangle
  \,=\, \langle k_1 k_2|V|k_3 k_4 \rangle \nonumber \\
 \!\!\!\!\!\!\!\!
  & & \quad  + \sum_{k_5, k_6}
    \frac{\langle k_1 k_2|V|k_5 k_6 \rangle \,Q(k_5,k_6) \, 
         \langle k_5 k_6 | G(\omega) |  k_3 k_4 \rangle}
        {\omega - e(k_5) - e(k_6)} \quad
\label{eqn:gmateq}
\end{eqnarray}
where $Q(k,k')=\theta(k-k_F)\theta(k'-k_F)$ is the Pauli exclusion operator
tp prevent two nucleons from scatterng into occupied states.
The single particle energy $e(k) = \frac{k^2}{2M_N} + U(k)$ contains the single-particle
potential $U(k)$ defined by
\begin{equation}
 U(k) = \!\sum_{k' \le k_F} {\rm Re\,} \langle k k'| G(e(k)+ e(k')) |k k'\rangle_A \ \ .
\label{eqn:self}
\end{equation}

The $G$ matrix is obtained by solving Eqs.(\ref{eqn:gmateq},\ref{eqn:self}) self-consistently, using
the lattice QCD $N\!N$ potential for $V$ and the lattice QCD nucleon mass for $M_N$.
Then the total energy $E$ is obtained from Eq.(\ref{eqn:gene}).
We employ the matrix inversion method~\cite{Haftel} with the continuous choice for $U(k>k_F)$~\cite{Song:1998zz}.
Because of the limitation for the lattice QCD potentials available at present,
we keep the partial-wave decomposition of the $G$-matrix only up to
 $^1{\rm S}_0$, $^3{\rm S}_1$, and $^3{\rm D}_1$ channels.

\begin{figure}[t]
\centering
\includegraphics[width=0.35\textwidth]{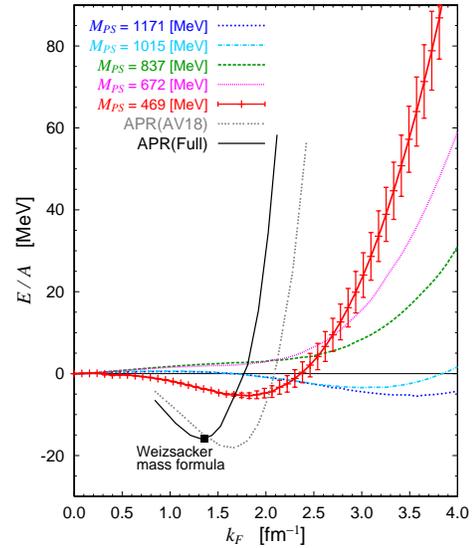}
\caption{Ground state energy per nucleon ($E/A$) for symmetric nuclear matter 
as a function of $k_F$ obtained by the BHF theory with the $N\!N$ potential from lattice QCD.
Filled square indicates the empirical saturation point, and
the curves labeled APR are taken from Ref.~\cite{Akmal:1998cf} with and without the 
phenomenological three nucleon force.
The error bars for the result at  $M_{\rm PS} \simeq 469$ MeV are 
the statistical uncertainties estimated by dividing the gauge configurations into two sets.}
\label{fig:nuclera_matter}
\end{figure} 

Figure.~\ref{fig:nuclera_matter} shows the ground state energy per nucleon ($E/A$)
for the symmetric nuclear matter ($Z=N=A/2$ with the proton number
$Z$, neutron number $N$ and the mass number $A=N+Z$)
as a function of $k_F$ for different $m_q$.
The most important feature of the symmetric nuclear matter 
 in the real world is its saturation property; i.e.,
  $E/A$ takes the minimum at normal nuclear matter density $\rho_0$.
The empirical saturation point from the Weizs\"acker mass formula
corresponds to $(k_F, E/A) \simeq (1.36 \ {\rm fm}^{-1}, -15.7\ {\rm MeV})$ as 
indicated  in Fig.~\ref{fig:nuclera_matter}. Also, we show the results of Ref.~\cite{Akmal:1998cf}
which employs the variational method with AV18 $N\!N$ potential
with and without phenomenological three-nucleon $N\!N\!N$ force.

Our result at the lightest quark mass ($M_{\rm PS}\simeq 469$ MeV) in Fig.~\ref{fig:nuclera_matter}
indicates that the symmetric nuclear matter becomes a self-bound system with 
a saturation point $(k_F, E/A) \simeq (1.83 \pm 0.16 \ {\rm fm}^{-1}, -5.4 \pm 0.5 \ {\rm MeV})$.
Here the errors are statistical uncertainties estimated 
by dividing the gauge configurations into two sets.
This is the first time that the nuclear force obtained from first principle lattice QCD simulations 
leads to the saturation in the symmetric nuclear matter.    
The saturation point, however, deviates from the empirical point primarily due to
heavy quark masses in our lattice simulation and the lack of $N\!N\!N$ force in our BHF calculation. 

\begin{figure}[t]
\centering
\includegraphics[width=0.35\textwidth]{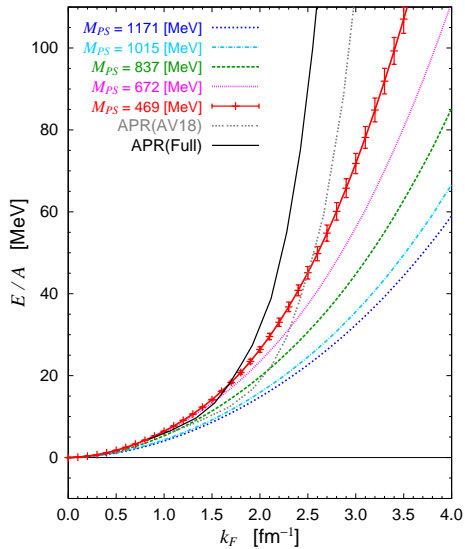}
\caption{Ground state energy per neutron for the pure neutron matter as a function of the Fermi momentum.
Details of the calculation are the same as Fig.~\ref{fig:nuclera_matter}.}
\label{fig:neutron_matter}
\end{figure} 

Also, we find nontrivial $m_q$ dependence of the EOS:
the saturation disappears at intermediate quark masses
($M_{\rm PS}\simeq 672, 837$ MeV) and appears again at the heavy quark mass region
($M_{\rm PS}\simeq 1015, 1171$ MeV).  This implies that the saturation 
originates from a subtle balance between short-range repulsion and the intermediate 
attraction of the nuclear force which have different $m_q$ dependence~\cite{Inoue:2011ai}.
The saturation points  for $M_{\rm PS}\simeq 1015, 1171$ MeV
 locate at more than 10$\rho_0$, so that
the effect of finite lattice spacing and the validity of BHF theory need to be carefully
checked for those cases.
Our restriction of the $N\!N$ interactions and the $G$ matrix 
to the $^1{\rm S}_0$ and $^3{\rm S}_1$-$^3{\rm D}_1$ channels may not be a bad approximation, 
since  the $S$ wave is known to be a dominant contribution and
 higher partial waves tend to cancel with each other near the physical saturation 
 point~\cite{Baldo:1991zz}. Nevertheless, we will include the
 results of on-going spin-orbit force calculation on the lattice~\cite{Murano:2013xxa}
 in the near future. In addition, we need to include the three-nucleon force from
 lattice QCD simulations~\cite{Doi:2011gq} to make the EOS at high density more realistic.

Figure.~\ref{fig:neutron_matter} shows  $E/A$ for the pure neutron matter ($N=A$) as a function 
of $k_F$. In this case, neutron matter is not self-bound due to large Fermi energy.
The pressure of the system is proportional to the slope of $E/A$ as
$P= \rho^2 \frac{\partial \, (E/A)}{\partial \rho} = \frac{k_F^4}{9\pi^2} \frac{\partial \,(E/A)}{\partial k_F}$;
Fig.~\ref{fig:neutron_matter} indicates that $P$ increases quite rapidly as $m_q$
decreases.

\begin{figure}[t]
\centering
\includegraphics[width=0.4\textwidth]{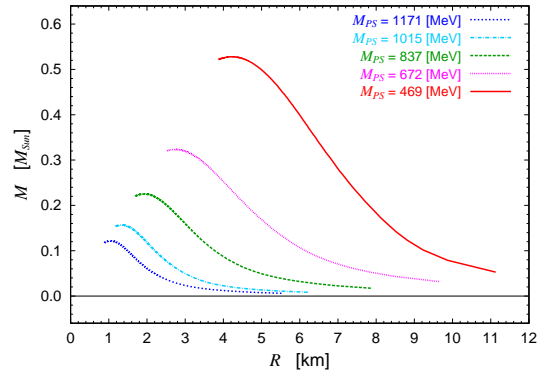}
\caption{Mass-radius relation of the neutron star.
Neutron-star matter consists of $n$, $p$, $e^-$, and $\mu^-$
with charge neutrality and chemical equilibrium.
EOS for the nucleons is obtained by an interpolation between Fig.~\ref{fig:nuclera_matter} 
and Fig.~\ref{fig:neutron_matter} under the parabolic approximation.
}
\label{fig:neutron_star}
\end{figure} 

To see the $m_q$ dependence of the EOS for neutron-star matter,
we calculate the mass ($M$) and the radius ($R$) of neutron stars for different $m_q$: 
The Tolman-Oppenheimer-Volkoff equation is solved 
 by using the EOS of neutron-star matter with
neutron, proton, electron and muon  under the charge neutrality and beta equilibrium.
For asymmetric nuclear matter, we use the parabolic approximation,
$\frac{E}{A}(\rho,x) = \frac{E_{Z=N}}{A}(\rho) + (1 - 2 x)^2 \epsilon_{\rm sym}(\rho)$
with the symmetry energy $\epsilon_{\rm sym}= E_{Z=0}/A - E_{Z=N}/A$ and the proton fraction $x = \rho_p/\rho$.
This is a standard interpolation between the symmetric nuclear matter and pure neutron matter.
The leptons (electrons and muons) are treated as the noninteracting Fermi gas. 
Solving the Tolman-Oppenheimer-Volkoff equation with a given value of central mass-energy density as an initial condition,
we obtain the $M$-$R$ relation for a spherically symmetric and nonrotating neutron star.

Shown in Fig.~\ref{fig:neutron_star} is the $M$-$R$ relation of the neutron star for different $m_q$.
As $m_q$ decreases, the $M$-$R$ curve shifts to the upper right direction, indicating
directly  the stiffening of our EOS.
The maximum mass of the neutron star ($M_{\rm max}$)
is found to be 0.53 times the solar mass ($M_{\odot}$) 
at the lightest $m_q$ corresponding to $M_{\rm PS}\simeq 469$ MeV and $M_N \simeq 1161$ MeV~\cite{radius}.
Such a maximum mass  is too small to account for the observed neutron stars
obviously due to the heavy $m_q$:
A naive extrapolation of $M_{\rm max}$ and the corresponding radius to $M_{\rm PS}=137$ MeV,
with a function $f(M_{\rm PS})=a/(M_{\rm PS}+b)+c$, for example,
gives  $M_{\rm max}=2.2 M_{\odot}$ and $R=12 $ km.
Although this is a crude estimate, it is encouraging for future quantitative studies \cite{crust}.

Throughout this Letter, the $NN$ potentials are taken from the 
zero-strangeness sector of the octet-baryon potentials obtained by lattice QCD simulations 
with flavor-$SU(3)$ symmetry.  In this case,
the vacuum polarization of the $s$ quark contributes equally to the $u$ and $d$ quarks,
though the valence quarks are restricted to only $u$ and $d$ quarks in the $NN$ sector. 
To be more realistic,  we need to consider explicit breaking of 
flavor-$SU(3)$ symmetry: 
Studies along this line on the lattice have been already started~\cite{Aoki:2012tk}
and results will be implemented in our future EOS calculation with the BHF theory.

To describe the nuclear matter and the neutron matter around the normal nuclear density,
it is sufficient to focus on the zero-strangeness sector. However, 
as the density exceeds a few times the normal nuclear density,
hyperons ($Y$) would start to appear in the ground state, which 
 is particularly relevant to the central core of neutron stars
(see e.g.~\cite{Masuda:2012kf} and references therein). Therefore,
construction of the EOS with $YN$ and $YY$ interactions is an important 
next step  in our approach.  
Since the hyperon forces are not well constrained by the experimental data,
the results of the lattice QCD simulations~\cite{Aoki:2012tk} are quite useful.
The three-baryon interactions~\cite{Doi:2011gq} with hyperons will  also be 
important to construct a realistic EOS to sustain recently discovered massive   neutron stars. 
The results of the present Letter provide a starting point of all these developments
in the near future.

\begin{acknowledgments}
We thank K.-I. Ishikawa and the PACS-CS group for providing their DDHMC/PHMC code~\cite{Aoki:2008sm},
and authors and maintainer of CPS++~\cite{CPS}, whose modified version is used in this Letter.
Numerical computations of this work have been carried out at Univ.~of Tsukuba supercomputer system (T2K).
This research is supported in part by Grant-in-Aid
for Scientific Research on Innovative Areas(No.2004:20105001, 20105003) and
for Scientific Research (B) 25287046, 24740146, (C) 23540321 and SPIRE (Strategic Program for Innovative REsearch).
T.H. was partially supported by RIKEN iTHES Project.
\end{acknowledgments}



\begin{thebibliography}{99}
\bibitem{radius}
  $E/A$ and its slope for $M_{\rm PS}$=469MeV in Fig.\ref{fig:neutron_matter}
  is already rather close to APR for $k_F < 1.5$ fm$^{-1}$.  
  This implies that small $M$ neutron star with the central density
  not far from $\rho_0$ have a radius of about 10 km.
  This is actually seen in Fig.\ref{fig:neutron_star}.
\end{thebibliography}
\end{document}